\documentstyle[12pt]{article}
  \date{}
  \title{$q$-deformed dynamics and  
   Josephson junction}
  \author{{Ramandeep S. Johal\thanks{e-mail: 
  rrai@lycos.com}}\\  
  {\it Department of Physics, Panjab University,}\\ 
  {\it Chandigarh -160 014, India. }} 
  \begin{document}
  \maketitle
  \def\be{\begin{equation}}
  \def\ee{\end{equation}}
  \def\ba{\begin{eqnarray}}
  \def\ea{\end{eqnarray}}
  \begin{abstract}
  We define a generalized rate equation for an observable in
  quantum mechanics, that involves a parameter $q$ and whose 
  limit $q\to 1$ gives the standard Heisenberg equation.
  The generalized rate equation is used to study 
  dynamics of current biased Josephson junction. 
  It is observed that this toy model incorporates
  diffraction like effects in the critical current.
  Physical interpretation for $q$ is provided which is 
  also shown to be $q$ deformation parameter. 
  \end{abstract}
  \newpage
  \baselineskip 24pt
  $q$-deformed algebras \cite{1,2} and their various  realizations have been
  subject
  of intense study in the past years.
  They provide a platform to study alternative 
  schemes of quantization \cite{3}  whereby issues like generalized
  statistics \cite{4} can be discussed. The effect of deformation has been
  studied also for different quantum mechanical toy models \cite{5a,5b}
  and these algebras are finding applications as dynamical symmetry
  algebras of various physical systems.
  Roughly speaking, $q$-deformed version of a theory
  corresponds to its  formulation on a lattice and $q$
  parameter may be interpreted as equivalent to lattice
  spacing \cite{6}.
  Appropriately, it is also interesting to observe that
  some theories can be viewed as intrinsically $q$-deformed, and
  thus use of $q$-deformed algebras appears naturally in
  certain contexts, like Barnett-Pegg theory of rotation angle \cite{7},
  nonextensive entropy within generalized statistical
  mechanics \cite{8}, Bloch electron problem \cite{9} and so on.

  Apart from the $q$-deformation of the algebraic structure
  for model physical systems,
  attempts have been made to $q$-generalize Schr\"{o}dinger 
  equation \cite{9c}, Dirac equation \cite{9d} as well as Heisenberg equation
  of 
  motion \cite{9e}. In this paper, we focus on the deformation of
  Heisenberg equation, which gives the rate of change
  of an observable $\hat{A}$ as
  \be
  \dot{\hat{A}}\equiv \frac{d \hat{A}}{dt} = \frac{1}{i\hbar}
  [\hat{A},\hat{H}].
  \label{heq}
  \ee
  Now a simple $q$-deformation of this equation,
  would be to make the rate of change proportional
  to modified commutator $[\hat{A},\hat{H}]_q =
  \hat{A}\hat{H} -q\hat{H}\hat{A}$. However, in this
  approach the hamiltonian is no longer constant of motion, 
  as pointed out in \cite{9e}.

  Our purpose here is to propose  a generalized rate equation
  for an observable in quantum mechanics.
  We apply it to study the dynamics of Josephson tunnel
  junction (JJ) \cite{10}.  The new definition involves a parameter
  which will be interpreted as $q$-deformation parameter and it will
  be discussed that it does lead to a physical effect.

  We first present a brief summary of the standard way of writing
  Heisenberg equation of motion for a JJ. Following \cite{11},
  the time dependent hamiltonian  of a current biased JJ is given by
  \be
  \hat{H} = \frac{1}{2C}\{ 2e\hat{n} + It\}^2 - E_J \cos{\hat{\phi}}.
  \label{ham}
  \ee
  $C$, $e$, $I$ respectively are the junction capacitance, electronic 
  charge,  current through the junction and $E_J$ is the Josephson coupling
  energy between two superconductors across the junction.
  $\hat{n}$ measures the number of cooper pairs transferred across
  the junction and $\hat{\phi}$ is the phase operator canonically conjugate
  to $\hat{n}$, in the sense that
  $[\hat{n}, \hat{\phi}] = -i$,
  which may be satisfied by the differential realization, 
  $\hat{n} = -i\frac{\partial}{\partial \phi}$.
  It may be remarked that $\hat{n}$ and $\hat{\phi}$  are hermitian
  operators and $\cos{\hat{\phi}}$ may be written in terms of unitary
  exponential operators as
  $\cos{\hat{\phi}} = (e^{i\hat{\phi}}+ e^{-i\hat{\phi}})/{2}$.   

  Now applying Eq. (\ref{heq}) for the case of $\hat{\phi}$ and $\hat{n}$
   (and using alongwith $[\hat{n},e^{i\hat{\phi}}]= e^{i\hat{\phi}}$),
   we get
  \be
  2e\dot{\hat{n}} = -I_J \sin{\hat{\phi}}, \quad
  (\hbar C/2e)\dot{\hat{\phi}} = 2e\hat{n} + It,
  \label{heq2}
  \ee
  where $I_J = 2eE_J/\hbar$ is the critical value of current,
  beyond which a finite voltage appears across the junction.

  Combining the two equations in (\ref{heq2}), we write
  \be
  (\hbar C/2e)\ddot{\hat{\phi}} + I_J \sin{\hat{\phi}}=I.
  \label{ddem}
  \ee
  In case, $E_J$ is quite larger
  than the electrostatic  charging energy $E_C = 2e^2/C$  
  (due to transfer of one cooper pair across the junction),
  we can replace the operators by their expectation
  values and so the "classical" equation of motion
  corresponding to Eq. (\ref{ddem}) is 
  \be
  (\hbar C/2e)\ddot{{\phi}} + I_J \sin{\phi}=I,
  \label{cem}
  \ee
  which implies motion of a classical particle 
  in  a washboard potential
  \be
  U(\phi) = -{2\over\hbar e}(I_J\cos{\phi} +I\phi).
  \label{pot}
  \ee
  In the absence of fluctuations, $\dot{\phi}=0$, and
  we have a minimum in potential energy, ${\partial U\over\partial \phi} =0$
  obtaining 
  \be
  I=I_J\sin\phi.
  \ee
  In the following, we  generalize the definition of
   rate of  change of  an operator with respect to time.
  Let us illustrate for an arbitrary observable $\hat{A}$.
  Define
  \be
  {\cal{D}}_t \hat{A} =  \frac{q^{-\hat{A}}}{{\ln}q}{d q^{\hat{A}}\over
  dt}.
  \label{def1}
  \ee
  The operator defined above is not equal to $d\hat{A}/dt$
  if $\hat{A}$ and  $d\hat{A}/dt$ do not commute.
  This can be seen by expanding $q^{\hat{A}}$ as power series.
   However, as $q\to 1$, the above
  operator approaches $d\hat{A}/dt$.  Also ${d q^{\hat{A}}}/{dt}$
  is given by the standard Heisenberg equation of motion
  \be
  \frac{d q^{\hat{A}}}{dt} = \frac{1}{i\hbar} [q^{\hat{A}},\hat{H}].
  \label{heqa}
  \ee
  Thus we obtain
  \be
  {\cal D}_t \hat{A} = \frac{q^{-\hat{A}}}{i\hbar {\ln}q}
                      [q^{\hat{A}},\hat{H}].
                     \label{fdeq} \ee
  So as $q\to 1$, the right hand side approaches
  $[\hat{A},\hat{H}]/i\hbar$ and in this sense, 
  Eq. (\ref{fdeq}) can be considered as $q$-generalization
  of the standard Heisenberg equation. We later on
  identify the parameter $q$ as the one appearing
  in $q$-deformation theory.

  To apply the new rate equation to Josephson junction,
  we alternately consider $\hat{A}$ to be
  operator $\hat{\phi}$ and $\hat{n}$.
  Also from now on we take $q= \exp(is)$ with $s$ real.
  As a consequence, we obtain
  \be
  (\hbar C/2e){\cal D}_t{\hat{\phi}} = 2e\hat{n} + It +es,
  \label{eom1}
  \ee
  and
  \be
  2e{\cal D}_t{\hat{n}} = -I_J\frac{\sin(s/2)}{s/2}{\sin}(\hat{\phi}-s/2).
  \label{eom2}
  \ee
  Compare these two equations with the original Heisenberg equations,
  Eq. (\ref{heq2}). Naturally, the above Eqs. reduce to the latter,
  as $s\to 0$ (or $q\to 1$). Following the similar arguments (as
  after Eq. (\ref{heq2})), we can identify the critical current
  of the modified problem as following
  \be
  I = I_J  \frac{{\sin}(s/2)}{s/2}{\sin}({\phi}-s/2).
  \label{cc}
  \ee
  There are two consequences of $s$ being not equal to zero; 1)
  there is phase shift, that in standard JJ may be produced by, for
  example, external magnetic field. It is this phase shift that
  accounts for interference in SQUID devices, where more than
  one JJs are coupled, 2) more remarkably, we see that
  the maximum critical current $I_J$ is  modified by a Fraunhoffer like
   diffraction term. This pattern is observed in junctions with rectangular
  geometry in the presence of applied magnetic field \cite{12}. This is a
  consequence of the finite width  of the junction and is
  analogous to the effect in optics. Comparing with the
  standard result, we  have $s = 2\pi \Phi/\phi_0$,
  where $\Phi$ is the effective 
  flux linked with the junction
  and $\phi_0 =\pi\hbar/e$.

  Next, we would like to interpret  parameter $q$
  as the one occuring in $q$-analysis. Note that
  using the realization $\hat{n} = -i\partial/\partial\phi$, we can show 
  \be
  q^{\hat{n}}e^{i\hat{\phi}}=q e^{i\hat{\phi}}q^{\hat{n}}.
  \label{qplc}
  \ee
  Now in the theory of $q$-calculus \cite{6}, the relation $\hat{X}\hat{Y}
  =q\hat{Y}\hat{X}$
  plays important role. Working in the eigenspace of $\hat{X}$,
  we have $\hat{X}\vert n\rangle = x\vert n\rangle$, where 
   $n$  labels the eigenstates.  On this space,
  $\hat{Y}$ acts like a shift operator, $\hat{Y}\vert n\rangle = \vert
  n+1\rangle.$
  For the present case, we  identify $X  =q^{\hat{n}}$
  and   $\vert n\rangle$ as  the pair-number difference states.
  Here $n=0,\pm 1,\pm 2,\cdots$. Note that $n$ counts the 
  number of cooper pairs transferred across the junction and
  the sign $\pm$ refers to the direction  (positive for
  transfer in one direction and negative for opposite 
  direction).   
   Also $\vert n\rangle$ 
  is the appropriate basis in which  the present hamiltonian
  Eq. (\ref{ham}), describes tunnelling of cooper pairs.
  Now the dual basis to the above is given by
  \be
  |\phi\rangle  = (2\pi)^{-1/2}
  \sum_{n=-\infty}^{\infty}e^{-i n\phi} \vert n\rangle, 
  \ee
  and the eigenvalue equation, $e^{i\hat{\phi}}\vert \phi\rangle =
  e^{i\phi}\vert \phi\rangle$,
  implies $e^{i\hat{\phi}}|n\rangle =|n +1\rangle $.
  Thus it is natural to
  identify  $ \hat{Y} =e^{i\hat{\phi}}$, satisfying Eq. (\ref{qplc}). 
  Also, $e^{-i\hat{\phi}}$ transfers the cooper pair in the opposite
  direction.
  It may be remarked that incorporating 
  negative eigenvalues of $\hat{n}$ is important to maintain
  unitarity of the shift operator $e^{i\hat{\phi}}$, and hence the
  hermiticity
  of operator  $\hat{\phi}$.

  In the above, we  used the standard JJ 
  hamiltonian with a generalized definition of
  rate of change of an observable and arrived at 
  generalized rate equations for the JJ. 
  It may be asked whether same generalized
  dynamics can be obtained by using a 
  modified or deformed hamiltonian in
  the standard Heisenberg equation.
   For the JJ case,
  it is easy to give such a hamiltonian as follows
  \be
  \hat{H}_s = \frac{1}{2C}\{ 2e(\hat{n}+{s/2}) + It \}^2 - {E_J}^{\prime}
   \cos(\hat{\phi}-{s/2}),
  \label{ham2}
  \ee
  where now the coupling energy becomes $s$ dependent as
  \be
  {E_J}^{\prime} = E_J {\sin{(s/2)}\over s/2}.
  \ee
  We note that $\hat{H}_s$ involves operators $\hat{n}$ and 
  $\hat{\phi}$ both shifted by $s/2$ (with opposite signs).
  To interpret the parameter $s$ in the modified hamiltonian,
  we consider the hamiltonian  for a charged particle moving on a ring 
  in the xy-plane
  \be
  \hat{H} = {{\hat{L}_z}^2\over 2M} - V_0\cos\hat{\phi},
  \ee
  where $M$ is the moment of inertia. Angular momentum
  operator $\hat{L}_z$ is the analogue of operator $\hat{n}$ in the case of
  JJ above and satisfies $[\hat{L}_z,\hat{\phi}] =i\hbar$. 
  Now the  standard equation of motion for $\hat{\phi}$ 
  is given by
  $\dot{\hat\phi} = \hat{L}_z/M.$
  The generalized rate of change  comes out to be 
  \be
  {\cal D}_t\hat\phi = (2\hat{L}_z +s)/2M.
  \ee
   It is clear that
  the following hamiltonian
  \be
  \hat{H}_s = {({\hat{L}_z+s/2})^2\over 2M} - V_0\cos\hat{\phi},
  \label{ham3}
  \ee
  when employed in the standard rate of change,
  gives the generalized dynamics.
  Now the deformed hamiltonian Eq. (\ref{ham3}) 
  is just the one for a charged particle moving  on a ring at whose
  center is magnetic flux proportional to $s$. Thus
  once again, we get the same interpretation of parameter
  $s$ as representing a fictitious flux. 

  Concluding, we have considered a generalized rate
  equation for an observable in quantum mechanics,
  which goes over to standard Heisenberg equation as 
  the parameter $q =\exp(is)$ goes to unity.
  To study its implications, we have applied it
  to current biased Josephson tunnel junction.
  We saw that this toy model can incorporate the 
  diffraction like effects of critical current for rectangular
  JJs.  This allows to interpret parameter $s$ as
  equivalent to magnetic flux through the junction.
  Also we were able to represent 
  $q$ as deformation parameter occuring in the theory 
  of $q$-analysis.   Finally, 
  we considered deformation of the system hamiltonian
  and again  could infer that $s$ represents a magnetic 
  flux threading the Josephson junction or the analogue system
  of charged particle on a ring.

  {\bf Acknowledgement}

  The author gratefully acknowledges Prof. K.N. Pathak for 
  discussion and encouragement. This work was supported
  by Department of Science and technology, India.
  
  \end{document}